\documentclass[prd,twocolumn,aps,showpacs]{revtex4}

\newcommand{\be}{\begin{equation}}
\newcommand{\ee}{\end{equation}}
\newcommand{\bn}{\begin{eqnarray}}
\newcommand{\en}{\end{eqnarray}}
\newcommand{\p}{\partial}
\def\ni{\noindent}

\begin{document} 

\title{New formulations for dual equivalent actions}
\author{E. M. C. Abreu$^{a}$, A. Calil$^{a}$, L. S. Grigorio$^{b}$, M. S. Guimaraes$^{b}$, and C. Wotzasek$^{b}$}
\affiliation{$\mbox{}^{a}$Departamento de F\'{\i}sica, Universidade Federal Rural do Rio de Janeiro\\
BR 465-07, 23851-180, Serop\'edica, Rio de Janeiro, Brazil.\\
{\sf E-mail: evertonabreu@ufrrj.br}\\
$\mbox{}^{b}$Instituto de F\'\i sica, Universidade Federal do Rio de
Janeiro, 21945, Rio de Janeiro, Brazil\\
{\sf E-mail: clovis@if.ufrj.br}\\
\today}


\begin{abstract}
New actions in $D=2$ and $D=3$ are proposed that are dual equivalent to known theories displaying well defined chirality and helicity, respectively, along with a new interpolating action that maps continuously through the original dualities.  The new chiral action in $D=2$ is a second-order theory displaying the chiral constraint dynamically while in $D=3$ the helicity constraint is imposed a la Siegel. The resulting theories introduce new versions of the Hull noton to take care of the symmetry aspects of the original models. The new interpolating formulation is then re-examined as a condensed phase for the discussion of duality under the light of the dual mechanisms -- Julia-Toulouse and Higgs -- establishing new interpolating actions in the dilute phase, according to these mechanisms.
\end{abstract}

\maketitle


\section{Introduction}

It is by now well-known that certain theories of anti-symmetric tensors,
among them gauge and scalar theories, admit description in terms of different
sets of potentials. This phenomenon is generally known as duality.
Therefore, the propagating modes written in terms of one of the potentials do have
description in terms of the dual potential, although the relation,
known as duality map, is in general non-local.

For systems with infinitely many degrees of freedom, dualities constitute
one of the very few analytic tools available for studying their non-perturbative
properties. Applications have been found in widely different physical settings,
ranging from spin models to quantum fields and string theories.
A duality transformation typically maps topological defects to fundamental fields, and vice
versa, and can therefore translate a non-perturbative problem to a solvable perturbative one.
The duality is mathematically an identity. In some cases the duality transformation can be carried out explicitly, mapping the partition function of one theory to that of another.
Therefore one can see exactly how the parameters and observables
of the two theories relate to each other.

A great deal of investigation on the nature of the duality has been carried out using the so called {\it master or parent action} technique \cite{hl}. In this approach an {\it intermediate} two-field action is postulated leading to the construction of the partition function of each of the dual aspects upon integration of one of the fields. This parent action is basically a reduced order action of one of them obtained through a Legendre transformation that introduces a second potential. The duality map is saw here as a sort of {\it discrete} symmetry transformation. What is of importance for the present investigation is that both ends of the duality map are frozen aspects of the transformation induced by the Legendre operation.

Other techniques to investigate duality have been introduced recently. For instance, the so-called gauging iterative {\it Noether dualization} method \cite{ainrw} has
been shown to thrive in establishing some dualities between models \cite{iw}.
This method is based on the traditional concept of a local lifting of a global symmetry and may be
realized by an iterative embedding of Noether counterterms.

An interesting physical aspect of the duality transformation is the mass generation problem. Recently, one of us and collaborators \cite{clovis}, have developed a systematic method for the studying mass generation and different aspects of duality, that embraces potentials of all ranks and dimensions.  It contains two basic elements; a point contact transformation and the soldering formalism that operates in the internal space \cite{bbg2}.  The mass generation is a consequence of the fusion of distinct fields and has been established in some special examples \cite{abw}. 

An alternative way of obtaining dual equivalence is the recent {\it dual embedding} method \cite{aon,amnot} that can avoid the introduction of infinite terms in the Hamiltonian of embedded non-commutative and non-Abelian theories.  This can be accomplished because the infinitesimal gauge generators are not deduced from previous unclear choices.
We should mention that this approach to embedding is not dependent on any undetermined constraint structure and also works for unconstrained systems.   This technique on the other hand only deals with the symplectic structure of the theory so that the embedding structure does not rely on any pre-existent constrained structure.
The method does not change the physical contents originally present in the theory computing the energy spectrum.  This technique follows Faddeev's suggestion \cite{FS} and is set up on a contemporary framework to handle noninvariant models, namely, the symplectic formalism \cite{FJ}.

A very powerful technique to study duality is the so called dual projection approach \cite{aw} which we strongly favor in this work. In this approach, the field equation differential operator (or its resolvent) is factorized into a pair of first-order operators (when possible) carrying, each of them, either the dynamical or the symmetry aspects of the original model.  However, once again, this technique only deals with the frozen aspects of the duality mapping. 

Another point of view, a midway path, advocated in this work, introduces a continuous mapping from one frozen aspect to the other through a new type of master action, in this way, continuously parameterizing the duality. Using this technique new actions to known dualities in D=2 and D=3 will be obtained. Extensions to higher dimensions are immediate. Also a link to different dualities related via condensation of topological defects either through the Higgs mechanism or via the Julia-Toulouse mechanism will be studied and established. This opens up a new door to the studies of duality by connecting models related to each other through phase transitions.

The paper is organized in such a way that in section 2, after a brief review of the Siegel theory and the noton concept, we introduce the new action and show its dual equivalence to the Floreanini-Jackiw (FJ) and Siegel actions.   After that, we study the modes of helicities present in the Proca theory in a 3D spacetime and analyze the duality of the three-dimensional Maxwell-Chern-Simons (MCS) model and show a new $D=3$ action using the Siegel method, demonstrating that the final spectrum comprises a Self-Dual particle and a $D=3$ noton.  In the final section we present our conclusions and perspectives and show, in particular, how the physical interpretation of the Julia-Toulouse and Higgs mechanisms may be incorporated into this new scheme leading to new representations of dualities in terms of an interpolating action, opening up a new window for the investigations on duality.

\section{The dual-decomposition and duality in 2D and 3D}

A great deal of investigation on the nature of the duality has been carried out on theories in 1+1 and in 2+1 dimensional Minkowski space. In the context of the latter the duality maps the fundamental fields of one theory to vortices of the other: the vortex lines can be understood as world lines of particles in the
dual theory. Three-dimensional space-times are characterized by the possibility of adding a gauge
invariant, non-conventional Chern-Simons term to the gauge field action. The resulting
theory is the MCS which (in Minkowski space-time) describes massive photons of a specific helicity.
A correspondence was first established by Deser and Jackiw \cite{dj} between the Maxwell-Chern-Simons 
theory and the so-called Self-Dual model \cite{tpn} using the parent action approach \cite{hl} leading to the paradigmatic duality in three dimensions. This mapping has since then been studied in more general settings such as, for example, to include the effects of external sources \cite{source}, to establish the correspondence between the partition functions for the massive Thirring model and the Maxwell-Chern-Simons (MCS) theories \cite{fs}, to include the effects of non-commutativity \cite{NC} and non-Abelian charges \cite{NA} as well as the breaking of Lorentz symmetry (in a 4D electrodynamics) \cite{cfjdual}.

The situation in $D=2$ Minkowski space-time is much less rich due to the absence of a topological structure such as the Chern-Simons term. However, due to its simplicity and the possibility to be used as a theoretical laboratory a great deal of investigation has also been performed in this dimension. A well known equivalence was established long ago between the fermionic Schwinger model and Abelian scalar field model with a vectorial interaction to the Maxwell field rendering mass to the "photon", a process known as Abelian bosonization. A much less known duality equivalence \cite{aw} was given between the chiral scalar models proposed by Siegel \cite{siegel} and by Floreanini and Jackiw \cite{fj2}.

What is quite interesting and appealing in these dualities is that in both dimensionalities there are Self-Dual type of models that are second-class constrained systems, first-order in time, describing a single chirality in $D=2$ \cite{FJ} and a single helicity in $D=3$ \cite{tpn}. In both cases the elimination of the degrees of freedom is a consequence of the presence of a self-dual constraint in the field equations of the theories. However, while there is a direct correspondence between these two models in different dimensions, i.e., between the Floreanini-Jackiw model and the SD-model, we find it compelling that such a direct correspondence does not exists for their associated dual theories. In $D=2$ the Siegel model which was shown to be the dual theory for the Floreanini-Jackiw model has its chiral constraint imposed by a non-dynamical Lagrange multiplier field, while in $D=3$ the Maxwell-Chern-Simons theory also presents its helicity constraint dynamically in the equations of motion. This suggests that there should be an action in $D=2$, dual to the FJ model, with a direct correspondence to the MCS model while one should expect that in $D=3$ the presence of another model, dual to the SD-model, with a direct correspondence to Siegel model. To explore and find such actions is the main point of this section.

In this section we shall use the dual projection approach to study the proposed new actions in two and three dimensions that are dual equivalent to these well known Self-Dual theories. We will show that it is possible to construct a Siegel-like theory in $D=3$ where the single helicity constraint is quadratically imposed over the Maxwell-Proca theory by using a Lagrange multiplier field.  The dual projection approach separates the dynamical sector of this model which presents the same dynamical contents as the SD model besides the existence of a Hull's \cite{hull} 3D noton-like sector. Similarly, we will show the existence of a second-order action in 2D, analogous to the MCS model, where the chiral constraint is imposed dynamically, as the dual projection approach will disclose.

\subsection{Duality and self-duality in 2D}

In order to consider the necessary features for the construction of the new action, in this section we first review the dual projection approach by studying the duality between Siegel and Floreanini-Jackiw models.  Then, the new second-order action with chiral dynamics will be examined.

The Siegel Lagrangian density for a left-mover scalar (lefton) is given by \cite{dgr},
\begin{equation}
\label{1order}
{\cal L} = \pi\,\dot{\phi}- \frac{{\phi'}^2}{2} -\frac{1}{2} \frac {\left(\pi-\lambda_{+}{\phi'}\right)^2}{1-\lambda_{+}}
-\frac{\lambda_{+}}{2}{\phi'}^2\,\, ,
\end{equation}
with an analogous result for the righton.
One can gauge-fix the value of the multiplier as $\lambda_{+} \rightarrow 1$ to reduce
it to its FJ form. The phase space of the model is correspondingly reduced to $\pi \rightarrow \phi'$.
The third term in (\ref{1order}) reduces to $(1-\lambda_{+}){\phi'}^2\rightarrow 0$
as $\lambda_{+}$ approaches its unit value.
This reduces the symmetry of the model, leaving behind
its dynamics described by a FJ action.
The above behavior suggests the following field redefinition,
\begin{eqnarray}
\label{theresult}
\phi = \varphi + \sigma;\;\qquad\;\qquad\;\pi = \varphi' - \sigma',
\end{eqnarray}
which is, in fact, a canonical transformation \cite{bg}.
 
The lefton $\phi$ is related to the FJ chiral mode $\varphi$ by the presence of a noton
$\sigma$. Such a decomposition immediately diagonalizes (\ref{1order}) as,
\begin{equation}
\label{III2}
{\cal L}  = ( {\varphi'}\,\dot{\varphi}\;-\;{\varphi'}^2)\;+\;
( -\;\sigma'\dot{\sigma}\;-\;\eta_+\,\sigma'^2)\;\;,
\end{equation}
with an analogous result for the righton, where 
\begin{equation}
\eta_{+} = \frac{1\;\;+\;\;\lambda_{+}}{1\;\;-\;\;\lambda_{+}} .
\end{equation} 
In this form, the chiral information is displayed by the FJ field ${\varphi}$ while the noton $\sigma$ carries the symmetry of the original model.
The reduction of the phase space is attained by letting the noton
$\sigma$ approach zero as the multiplier $\eta_+$ diverges.  This eliminates the symmetry carrying sector leaving behind only the FJ mode.  Hence, all the dynamical information of the theory is carried by the FJ mode.   We can say thereby that the Siegel and the FJ actions are dynamically equivalent. We may also ask for the meaning of the gauge fixing $\lambda_{+} \rightarrow -1$. This will be clarified below.

\bigskip

As we said previously, we are looking for a new action that is dynamically equivalent to the FJ one.
With this idea in mind we introduce such action as
\be \label{1.0}
{\cal L}_{2D} = {1\over2}{\phi'}\dot{\phi}\,+\,{1\over2}{\dot{\phi}}^2\,\, .
\ee
The next step is to write (\ref{1.0}) in first order which is,
\be \label{1.1}
{\cal L}_{2D} = \pi \dot{\phi}\,-\,{1\over2}(\pi\,- {1\over2}{\phi'})^2\,\,.
\ee
Performing the convenient canonical transformations,
\be
\label{1.2}
\phi = \varphi_+ + \varphi_-\,\,;\qquad\;\;\;\pi = {1\over2}{\varphi'}_+ \,-\, {1\over2}{\varphi'}_-\,\,,
\ee
and substituting them back in (\ref{1.1}) we have that,
\be
{\cal L}_{2D}\,=\,-\,{1\over2}{\varphi'}_{-}{\dot{\varphi}}_{-}\,+\,\left({1\over2}{\varphi'}_{+}\,{\dot{\varphi}}_{+}\,-\,{1\over2}{\varphi'}_{+}^{2}\right)\,\,.
\ee
We can see clearly that the last two terms in the action corresponds to a FJ chiral boson with the appropriated chirality.  It is also easy to see that the first term in ${\cal L}_{2D}$ describes a zero Hamiltonian system, which is a well known problem.  In this kind of system we have to work in a reduced phase-space \cite{zerohamiltonian,ht}, which is equivalent to describe the system in terms of constants of the motion \cite{ht}.  This fact leaves no generator of dynamics in the reduced phase-space.
The fact that this term represents no-dynamics show us that ${\cal L}_{2D}$ is dynamically equivalent to the FJ action.  Hence, it is clear that the action (\ref{1.0}) is also dynamically equivalent (dual) to the Siegel action and we can write,
\be
{\cal L}_{2D}\:\sim\: {\cal L}_{FJ}\:\sim\: {\cal L}_{Siegel}\,\,.
\ee

Now it comes a new twist. The new action (\ref{1.0}) is readily obtained from the Siegel action (\ref{III2}) by the gauge-fixing $\lambda_{+} \rightarrow -1$ discussed above and a trivial scaling of the fields.  In this sense therefore both the Floreanini-Jackiw action and the new action being introduced here derive from the Siegel model by appropriated gauge fixings. This observation allows us to look at the Siegel model as a new form of master action continuously interpolating between the two frozen dual aspects of the chiral boson theory. Such a concept is not restricted by dimensional considerations. We shall explore this new possibility next in order to construct a master action interpolating between the well known dualities in D=3.

\subsection{An alternative dual model in D=3}

Before discussing the new Self-Dual action let us quickly review the dual projection technique in the massive 3-D electrodynamics. Concerning dimensionality, the Proca action contains two massive degrees-of-freedom and parity is not broken.  So, let us write,
\be \label{2.0}
{\cal L}_{Proca}\,=\,-\,{1\over4}\,F_{\mu\nu}^2\,+\,{m^2\over2}\,A_\mu^2\;\;.
\ee
Here, differently from the $D=2$ case, the decomposition is totally covariant.  In this procedure we will use two well known identities,
\be \label{2.2}
(i) \quad \pm {1\over2}\,B_\mu^2\,=\,\pi_\mu\,B^\mu\,\mp\,{1\over2}\pi_\mu^2\;\;,
\ee
\be \label{2.3}
(ii) \quad {1\over2}\,F_{\mu\nu}^2\,=\,(\epsilon_{\mu\alpha\beta}\,\p^{\alpha}A^\beta)^2\;\;,
\ee
where $B_\mu$ is any physical quantity and $F_{\mu\nu}$ is the Maxwell stress tensor.

The identity (\ref{2.2}) is essentially a Legendre transformation as well as a first order reduction.  Substituting firstly Eq. (\ref{2.3}) and then Eq. (\ref{2.2}), the Proca action, Eq. (\ref{2.0}), in first-order formulation can be written as,
\bn \label{2.4}
{\cal L}_{Proca}\,&=&\,-\,{1\over2}\,(\epsilon_{\mu\alpha\beta}\,\p^{\alpha}A^\beta)^2\,+\,{m^2\over2}\,A_\mu^2  \\
&=&\pi^{\mu}\,(\epsilon_{\mu\alpha\beta}\,\p^{\alpha}A^\beta)\,+\,{m^2\over2}\,A_\mu^2\,+\,{1\over2}\pi_{\mu}^2\,\,. \nonumber
\en
We can perform the dual-decomposition (canonical transformation) through a ${\pi / 4}$ rotation in the following space, i.e., $(\pi_\mu,A_\mu) \rightarrow (f_\mu,g_\mu)$, we can write that,
\bn \label{2.5}
A_\mu = f_\mu\,+\,g_\mu \qquad \mbox{and} \qquad
\pi_\mu = m\,(f_\mu\,-\,g_\mu)\,\,.
\en
Hence, using (\ref{2.5}) in (\ref{2.4}), we have,
\bn \label{2.55}
{\cal L}_{Proca} &=&(m\,\epsilon_{\mu\alpha\beta}\,f^\mu\,\p^\alpha\,f^\beta\,+\,m^2\,f_\mu^2)\, \nonumber \\
&+&\,(- m\,\epsilon_{\mu\alpha\beta}\,g^\mu\,\p^\alpha\,g^\beta\,+\,m^2\,g_\mu^2)\,\,,
\en
this diagonalized action is an expected result since the Self-Dual actions that compose the Proca theory have opposite self-dual degrees-of-freedom each one.  This dual-decomposition is analogous to that of the $D=2$ case, it differs only in the case that now the excitations are massive.

We are now in position to discuss the dual equivalence between the Maxwell-Chern-Symons (MCS) and the Self-Dual model in this dual-decomposition context which we are dealing with.
Let us write the MCS action and substitute the equations (\ref{2.2}) and (\ref{2.3}) in the same way as we did before in order to obtain a first-order formulation,
\bn \label{2.11}
{\cal L}_{MCS}\,(A_{\mu})&=&-\,{1\over4}\,F_{\mu\nu}^2\,-\,{m\over2}\,A^\mu\,\epsilon_{\mu\alpha\beta}\,\p^\alpha A^\beta \\
&=&(\pi^\mu\,-\,{m\over 2}\,A^\mu)\,(\epsilon_{\mu\alpha\beta}\,\p^\alpha\,A^\beta)\,+\,{1\over2}\pi_\mu^2\,\,. \nonumber
\en
\ni Let us conveniently define
\be
\Pi_\mu\,=\,\pi_\mu\,-\,{m\over 2}\,A_\mu \,\,,
\ee
\ni and substituting it back in (\ref{2.11}) we have that,
\be \label{2.111}
{\cal L}_{MCS}\,=\,\Pi^\mu\,(\epsilon_{\mu\alpha\beta}\,\p^\alpha\,A^\beta)\,+\,{1\over2}(\Pi_\mu\,+\,{m\over 2}\,A_\mu)^2\,\,.
\ee
Making canonical transformations similar to those in Eqs. (\ref{2.5}) we see that the term,
\be \label{2.13}
\Pi_\mu\,+\,{m\over 2}\,A_\mu\,=\,m\,f_\mu\,\,.
\ee
\ni Hence, we can write Eq. (\ref{2.111}) as
\bn \label{2.14}
{\cal L}_{MCS}\,&=&\,(m^2\,f_\mu^2\,+\,m\,f^\mu\,\epsilon_{\mu\alpha\beta}\,\p^\alpha\,f^\beta)\, \nonumber \\
&+&\,(-\,m\,g^\mu\,\epsilon_{\mu\alpha\beta}\,\p^\alpha\,g^\beta)\,\,,
\en
which, after an appropriated scaling, was naturally separated in two terms.  The first and the second one are $f$ and $g$ dependent respectively.
This proves that in fact the Self-Dual action ${\cal L}^+ (f)$ (the first term in Eq. (\ref{2.14})) is dynamically equivalent to the MCS action Eq. (\ref{2.11}).

Since it is clear that in the dual-decomposition introduced in (\ref{2.14}) the self-dual dynamics is carried by the $f$ component, one can ask about the r\^ole played by the $g$ component in this context.  We believe that $f$ carries the dynamics of the theory while $g$ is responsible for the gauge symmetry of the MCS model.

As ${\cal L}^+\,(f)$ in (\ref{2.14}) is not gauge invariant let us propose the following transformation
\be \label{2.18.a}
A_\mu \quad \rightarrow \quad A_\mu\,+\,\p_\mu\,\eta
\ee
\ni in the MCS model so that the dual-decomposition (\ref{2.5}) reads
\be \label{2.18.b}
f_\mu\quad\rightarrow\quad f_\mu
\ee
\be \label{2.18.c}
g_\mu\quad\rightarrow\quad g_\mu\,+\,\p_\mu\,\eta
\ee
and, not considering total derivatives, we have that the second term in Eq. (\ref{2.14}) is invariant.

In this way, we can say that the dual-decomposition introduced a new and interesting interpretation of the dual equivalence between the Self-Dual and $MCS$ models with a clear definition about the dynamical and the symmetry sectors.

Exploring the analogy between the $D=2$ and $D=3$ models, we can write the following suggestive correspondences,
\be
{\cal L}_{FJ}\,=\,\dot{\phi}\,\phi'\,-\,{\phi'}^2\quad\rightarrow\quad m\,f^\mu\,\epsilon_{\mu\alpha\beta}\,\p^\alpha\,f^\beta\,+\,m^2\, f_\mu^2
\ee
\be
{\cal L}_{D=2}\,=\,\dot{\phi}\,\phi'\,-\,{\dot{\phi}}^2\quad\rightarrow m\,A^\mu\,\epsilon_{\mu\alpha\beta}\,\p^\alpha\,A^\beta\,-\,{1\over4}F_{\mu\nu}^2
\ee
so that the results above would deal with the equivalence between the chiral bosons formulations and the second-order Self-Dual and MCS theories respectively.  With these analogies discussed above, we can see clearly that the considerations about the dynamics and symmetry properties in one dimension can be brought to the next one.
In the sequel, we will construct a new action in $D=3$ interpolating between these known dualities in clear analogy to the 2D Siegel model.

\bigskip
\bigskip

Let us now consider the question that shows up naturally about the existence of a different and new model equivalent to the SD and MCS models and that appears due to the self-dual constraint imposition.  From Siegel theory discussed previously we know that this constraint must be quadratic to prevent the Lagrange multiplier from getting dynamics.   In our case the imposition can be obtained by constraining the Proca model in order to eliminate one of the self-dual propagation modes. 
So, let us write analogously, in $D=3$, but now we will impose the self-dual constraint as,
\be \label{3.23}
{\cal L}_{D=3}^{(\pm)}\,=\,{\cal L}_{Proca}\,+\,\lambda\,(m\,A_{\mu}\,\pm\,\epsilon_{\mu\alpha\beta}\,\p^{\alpha}\,A^{\beta})^2\,\,,
\ee
in order to prove that,
\be \label{3.24}
{\cal L}_{D=3}\,\sim\,{\cal L}_{MCS}\,\sim\,{\cal L}_{SD}\,\,,
\ee
and in this way there is a new action in $D=3$ equivalent to the Self-Dual and MCS one.
To go on in this objective let us develop the Eq. (\ref{3.23}) so that,
\bn \label{3.25}
& &{\cal L}_{D=3}^{(\pm)} \nonumber \\
&=&\,-\,{1\over4}\,F^2_{\mu\nu}\,+\,{m^2\over2}\,A^2_\mu\,+\,{\lambda\over2}\,\left(m\,A_{\mu}\,\pm\,\epsilon_{\mu\alpha\beta}\,\p^{\alpha}\,A^{\beta}\right)^2 \nonumber \\
&=&-\,{1\over4}\,F^2_{\mu\nu}\,+\,{m^2\over2}\,A^2_\mu\, \\
&+&\,{\lambda\over2}\,\left[\,m^2\,A^2_\mu\,+\,\left(\epsilon_{\mu\alpha\beta}\,\p^{\alpha}\,A^{\beta}\right)^2\,\pm\,2\,m\,A^\mu\,\epsilon_{\mu\alpha\beta}\,\p^{\alpha}\,A^{\beta}\,\right] \nonumber \\
&=& -\,{1\over4}\,(1\,-\,\lambda)\,F^2_{\mu\nu}\,+\,{m^2\over2}\,(1\,+\,\lambda)\,A^2_\mu\, \nonumber \\
&\pm&\,\lambda\,m\,A^\mu\,\epsilon_{\mu\alpha\beta}\,\p^{\alpha}\,A^{\beta}. \nonumber 
\en
Its first-order form is given, have after a little algebra, by
\bn \label{3.251}
{\cal L}_{D=3}^{(\pm)}&=&\,\pi^\mu\,\epsilon_{\mu\alpha\beta}\,\p^{\alpha} A^{\beta}\,+\,\frac{1}{2(1\,-\,\lambda)}\,\pi_\mu^2  \nonumber \\
&\,+\,&\frac{m^2}{2(1\,-\,\lambda)}\,(1\,-\,\lambda^2)\,A_\mu^2\,  
\pm\,\lambda\,m\,A^\mu\,\epsilon_{\mu\alpha\beta}\,\p^{\alpha}\,A^{\beta} \nonumber \\
&=&\Pi^\mu\,(\epsilon_{\mu\alpha\beta}\,\p^{\alpha}\,A^{\beta})\,  \\
&+&\,\frac{1}{2(1\,-\,\lambda)}\left[\,\Pi_\mu^2\,+\,m^2\,A_\mu^2\mp\,2\,\lambda\,m\,A_\mu\,\Pi^\mu\,\right]\;\;, \nonumber 
\en
where, now we have,
\be
\Pi_\mu\quad =\quad\pi_\mu\,\pm\,\lambda\,m\,A_\mu\;\;.
\ee
Using the dual projection as given in Eq. (\ref{2.5}), we have that,
\bn
& &{\cal L}_{D=3}\,=\,m\,(f^\mu\,\epsilon_{\mu\alpha\beta}\,\p^{\alpha}\,f^{\beta}\,-\,g^\mu\,\epsilon_{\mu\alpha\beta}\,\p^{\alpha}\,g^{\beta})\, \nonumber \\
&+&\,\frac{m^2}{(1\,-\,\lambda)}\left[f_\mu^2\,+\,g_\mu^2\mp\lambda\,(f_\mu^2\,-\,g_\mu^2)\right]\;\;.
\en
Now we have two possibilities depending on the relative self-dual mode to be eliminated by the constraint,
\bn \label{3.26a}
& &{\cal L}^{(+)}_{D=3}\,=\,\left(m\,f^\mu\,\epsilon_{\mu\alpha\beta}\,\p^{\alpha}\,f^{\beta}\,+\,m^2f_\mu^2 \right)\, \nonumber \\
&+&\,\left(-\,m\,g^\mu\,\epsilon_{\mu\alpha\beta}\,\p^{\alpha}\,g^{\beta}\,+\,m^2\,{(1\,+\,\lambda)\over(1\,-\,\lambda)}\,g_\mu^2\right)
\en
or
\bn \label{3.26b}
& &{\cal L}^{(-)}_{D=3}\,=\,\left(m\,f^\mu\,\epsilon_{\mu\alpha\beta}\,\p^{\alpha}\,f^{\beta}\,+\,m^2\,{(1\,+\,\lambda)\over(1\,-\,\lambda)}\,f_\mu^2\right)\, \nonumber \\
& &\qquad \qquad +\,\left(-\,m\,g^\mu\,\epsilon_{\mu\alpha\beta}\,\p^{\alpha}\,g^{\beta}\,+\,m^2\,g_\mu^2\right)\;\;,
\en
and in both actions above we can see a Self-Dual term.  

In the last section we analyze the spectra obtained and consider that the self-dual field would be responsible for the dynamics of the system.  The other term would carry the symmetry.
Hence, we understand that the same approach can be used here and the following relation can be established ,
\be
\label{3.27}
{\cal L}_{D=3}^{(\pm)}\,\sim\,{\cal L}_{SD}^{(\pm)}\,+\,{\cal L}_{Noton}\,\,.
\ee

\noindent From our experience in $D=2$, it is clear that the second term is the analogous to the symmetry term present in the Siegel action, i.e., it represents the $D=3$ version of Hull noton, which is new in the literature.

In other words, in $D=2$ we have,
\be \label{3.28}
{\cal L}_{Noton}^{D=2}\,=\,\dot{\varphi}\varphi'\,-\,\eta_2\,{\varphi'}^2 \qquad ; 
\qquad \eta_2\,=\,{(1\,+\,\lambda)\over(1\,-\,\lambda)}\,\,,
\ee
where $\eta_2$ is a Lagrange multiplier which eliminates the dynamics of this fields.   In $D=3$, we have analogously,
\be \label{3.281}
{\cal L}_{Noton}^{D=3}\,=\,\left(m\,\epsilon_{\mu\alpha\beta}\,h^\mu\,\p^{\alpha}\,h^{\beta}\,+\,m^2\,\eta_3\,h_\mu^2\right)\, \,\,;
\ee
where $\eta_3=\frac{1\,+\,\lambda}{1\,-\,\lambda}$.   

Hence, we can conclude that also in $D=3$ the primary function of the Hull field is to carry a representation of the symmetry of the system, i.e., that this field is present only in order to preserve the covariance of the theory but does not interfere in its dynamics.

\bigskip

Let us now consider the general case where a Chern-Simons term is added to the Proca model in order to discuss the 
effectiveness of the dual projection approach.  The general action is, 
\be \label{3.31}
{\cal L}\,=\,-\,{1\over4}\,F_{\mu\nu}^2\,+\,c_1\,m\,A^\mu\,\epsilon_{\mu\alpha\beta}\,\p^{\alpha}\,A^{\beta}\,+\,{m^2\over2}\,c_2\,A_\mu^2\,\,,
\ee
and in first order is,
\be \label{3.311}
{\cal L}\,=\,\Pi^{\mu}\,\epsilon_{\mu\alpha\beta}\,\p^{\alpha}\,A^{\beta}\,+\,{1\over2}\,(\Pi_{\mu}\,-\,c_1\,m\,A_{\mu})^2\,+\,{m^2\over2}\,c_2\,A_\mu^2\,\,,
\ee
where $\Pi^{\mu}=\pi^{\mu}\,+\,c_1\,m\,A^{\mu}$.

To make an analogous analysis as before we have to perform a dual projection, see (\ref{2.5}), the final action is,
\bn \label{3.32}
& &{\cal L}\,=\,\left[m\,f^\mu\,\epsilon_{\mu\alpha\beta}\,\p^{\alpha}\,f^{\beta}\,+\,(1\,-\,c_1)\,m^2\,f_\mu^2\right] \nonumber \\
&\,+\,&\left[-\,m\,g^\mu\,\epsilon_{\mu\alpha\beta}\,\p^{\alpha}\,g^{\beta}\,+\,(1\,+\,c_1)\,m^2\,g_\mu^2\right]\,\,, 
\en
where we fixed
\be
c_2 \,=\,1\,-\,c_1^2
\ee
as the condition to eliminate the cross terms.  We can see that if $c_1 \not= 0,\pm 1$ we have that the spectrum of the Proca MCS action is composed of two Self-Dual actions.  At the same time, if we choose $c_2=0\,\Rightarrow\,c_1=\pm 1$ or $c_1=0\,\Rightarrow\,c_2=1$, it is easy to see that we will have the previous cases again.

The important point to stress at the end of this section is that, as in the D=2 discussion, here too our new action (\ref{3.23}) plays the role of a master action producing the frozen dualities, i.e., the Maxwell-Chern-Simons and the Self-Dual actions, as special cases by gauge-fixing the Lagrange multiplier as $\lambda \to \mp 1$, respectively. Of course such a feature is not restricted by dimensional considerations and may be found for the massive self-dualities in all odd-dimensional spaces. It would be interesting to investigate if this new master action could be applied to other dualities besides the self-dual cases. That this seems to be the case will be indicated by the discussion in the next Section where a physical picture involving the dual instruments for charge condensation, i.e., the Higgs and the Julia-Toulouse mechanisms will be discussed.

\section{Conclusions and perspectives}

In this work we used the concept of dual-decomposition or dual-projection in order to 
introduce new actions in two and three dimensions that are dual equivalent to well known theories.  This was accomplished through the 
analysis of the energy spectrum of each final theory obtained.

In $D=2$ we introduce a new ``chiral boson" theory with a quadratic time derivative term which is different from the FJ quadratic differential space term.  We show that this action comprises a zero Hamiltonian term and a FJ particle.  While the former brings no-dynamics, the FJ part is responsible for the physical significance of the theory in phase-space.  Hence, the dynamical equivalence of this new action to the FJ one is plainly obvious.

In $D=3$ we review both that the spectrum of the Proca model is composed by two Self-Dual actions of opposite helicities and that the spectrum of the MCS model contains a Self-Dual particle, carrying the dynamics of the theory, and a term that is responsible by the gauge symmetry of the theory. Again, from the point of view of the dual projection, the dynamical equivalence of the MCS and the SD models are well understood.

With this idea in mind in $D=3$, we extended the Siegel approach to D=3 in order to construct a new action that would be dual equivalent to both the MCS and to the Self-Dual action and would contain both as particular
gauge-fixing points.  To demonstrate this we imposed a quadratic self-dual constraint over the 3D Proca model in order to eliminate one of the helicities.  The remaining model contains a Self-Dual mode and a totally new 3D version of the Hull's noton.  Again in $D=3$, such a noton is responsible for the symmetry of the theory.
As a final result we discuss the spectrum of a general Proca-MCS theory, imposing conditions in the coefficients in order to obtain two Self-Dual models.

Although these are all new and relevant results, it is quite evident from the above construction the emergence of a new structure containing all these results as particular cases and having the conventional duality as frozen aspects of a continuous transformation parameterized by the Lagrange multiplier field. As so one can appreciate the well know dualities in D=3 as gauge-fixing aspects of this new theory that contains them all. Before that, in D=2, after introducing a new action, dual in the traditional sense to the FJ model, the chiral boson model proposed by Siegel was reinterpreted as such a master theory in that dimension. Extensions to other dimensions are straightforward.

It is an interesting exercise to look at this new interpolating action from the perspective of the Julia-Toulouse versus Higgs mechanism. As we will see, this will lead to another set of interpolating actions with their frozen aspects being related either by a Higgs or a Julia-Toulouse mechanism. The Julia-Toulouse theory is a mechanism that takes in consideration the change in the physics of a relativistic system when it undergoes a phase transition due to the condensation of topological defects. It has been shown by Quevedo and Trugenberger to be the exact dual of the Higgs mechanism. As so, when two models are related to each other by a Higgs condensation, their duals are shown to be related by the corresponding Julia-Toulouse mechanism. Recently some of us and collaborators \cite{santiago1} have shown that such a relationship can also be extended to include quantum radiative corrections. As so such corrections may be alternatively computed via Julia-Toulouse mechanism. We can make a little exercise here trying to guess what such information would lead us from the master duality introduced in the preceding section. We will argue below that from (\ref{3.23}) we get the interpolating action for the $D=3$ scalar-vector duality 

\bn \label{scalar-vetor}
& &{\cal L} = -{1\over4}F^2_{\mu\nu}+{1\over2}\partial_\mu^2\phi +{\lambda\over2}\left(\partial_\mu\phi\,+\,\epsilon_{\mu\alpha\beta}\,\p^{\alpha}\,A^{\beta}\right)^2 , 
\en
when considering the transition back from a condensed to a diluted phase \cite{santiago1}. That this should be so is already apparent if one considers the MCS-SD duality inherent in the following formulae
\be
-{1\over4}F^2_{\mu\nu} + m A_\mu \epsilon^{\mu\nu\lambda}\partial_\nu A_\lambda \leftrightarrow
A_\mu^2 - {1\over m} A_\mu \epsilon^{\mu\nu\lambda}\partial_\nu A_\lambda \, .
\ee
Now, if we take the dilute limit $m\rightarrow 0$ in the SD sector then we get the representation of the vector field in terms of the scalar field, $m\,A_\mu \rightarrow \partial_\mu\phi$, coming from the dilute phase constraint $\epsilon^{\mu\nu\lambda}\partial_\nu A_\lambda \rightarrow 0$. This {\it jump of rank} phenomenon is a clear signature of the Julia-Toulouse mechanism \cite{QT}. On the MCS side of the duality the dilute limit leads to the Maxwell form directly. Introducing this limit into the Master action (\ref{3.23}) leads us directly to (\ref{scalar-vetor}) as proposed. Gauge-fixing $\lambda \rightarrow \pm 1$ leads to the two sides of the well known vector-scalar duality in D=3. Therefore, from the master action for MCS-SD duality we have arrived at a new master action (\ref{scalar-vetor}) related to each other via Higg/Julia-Toulouse dual mechanisms.

\section{Acknowledgments}

The authors would like to thank Funda\c{c}\~ao de Amparo \`a Pesquisa do Estado do Rio de Janeiro (FAPERJ) and Conselho Nacional de Desenvolvimento Cient\' ifico e Tecnol\'ogico (CNPq) (Brazilian agencies) for financial support.

\end{document}